# An On-Orbit CubeSat Centrifuge for Asteroid Science and Exploration

Jekan Thangavelautham
Space and Terrestrial Robotic
Exploration Laboratory
College of Engineering
University of Arizona
jekan@arizona.edu

Erik Asphaug
Lunar and Planetary Laboratory
College of Science
University of Arizona
asphaug@lpl.arizona.edu

Stephen Schwartz
Lunar and Planetary Laboratory
College of Science
University of Arizona
srs51@email.arizona.edu

*Abstract*— There are thousands of asteroids in near-Earth space and millions expected in the Main Belt. They are diverse in their physical properties and compositions. They are also time capsules of the early Solar System making them valuable for planetary science, and are strategic for resource mining, planetary defense/security and as interplanetary depots. But we lack direct knowledge of the geophysical behavior of an asteroid surface under milligravity conditions, and therefore landing on an asteroid and manipulating its surface material remains a daunting challenge.

Towards this goal we are putting forth plans for a 12U CubeSat that will be in Low Earth Orbit and that will operate as a spinning centrifuge on-orbit. In this paper, we will present an overview of the systems engineering and instrumentation design on the spacecraft. Parts of this 12U CubeSat will contain a laboratory that will recreate asteroid surface conditions by containing crushed meteorite. The laboratory will spin at 1 to 2 RPM during the primary mission to simulate surface conditions of asteroids 2 km and smaller, followed by an extended mission where the spacecraft will spin at even higher RPM. The result is a bed of realistic regolith, the environment that landers and diggers and maybe astronauts will interact with. The CubeSat is configured with cameras, lasers, actuators and small mechanical instruments to both observe and manipulate the regolith at low simulated gravity conditions. A series of experiments will measure the general behavior, internal friction, adhesion, dilatancy, coefficients of restitution and other parameters that can feed into asteroid surface dynamics simulations. Effective gravity can be varied, and external mechanical forces can be applied.

These centrifuge facilities in space will require significantly less resources and budget to maintain, operating in LEO, compared to the voyages to deep space. This means we can maintain a persistent presence in the relevant deep space environment without having to go there. Having asteroid-like centrifuges in LEO would serve the important tactical goal of preparing and maintaining readiness, even when missions are delayed or individual programs get cancelled.

TABLE OF CONTENTS

## 1. INTRODUCTION

The Planetary Science Decadal Survey highlights the need to explore small bodies to determine their origins, composition and surface-processes [1]. The scientific and practical importance of asteroids is represented by 7 current missions in various phases: Dawn [2], Psyche [3], Lucy [4], Hayabusa2 [5], OSIRIS-REx [6], DART/Hera [7-8] and MMX [9]. Still, landing on a small body is a complex challenge (e.g. the fate of Rosetta's Philae [10]) and so is manipulating the surface material in any way. Our ignorance about asteroid surface geomechanics is a strategic knowledge gap that slows down further exploration, making landings either risky, or highly limited, or both.

We propose a faster, cheaper, highly focused approach: an enhanced version of the Asteroid Origins Satellite 1 [11-14] called AOSAT+ to develop an asteroid surface laboratory lasting 1.5 yrs. or more in LEO. Experiments will play out for hours in this $10^{-5}$ to $10^{-3}$ g research centrifuge simulating small body surfaces (Figure 1). The system will permit days of nonstop experiments, using an array of actuators, cameras and sensors to create geophysical situations (impacts, gas jets, landslides) to study and interpret.

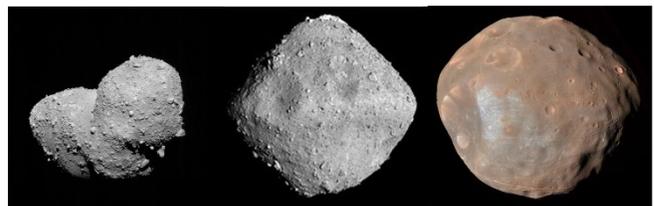

Figure 1: AOSAT+ will simulate the surface gravity of Itokawa, Ryugu and Phobos/Deimos.

Data will be downlinked each month, and new experiments uplinked, to enable hypothesis testing in real regolith materials under asteroid-like gravity, and to learn how to explore further in near-Earth space and beyond.

Experiments are needed at this juncture, and hence the requirement for a persistent link to an asteroid laboratory. If we wait to learn sequentially, one landed mission at a time, it will take decades and billions of dollars to move towards reliable robotic telepresence and human activity on asteroids.

That was to be the function and purpose of the Asteroid Return Mission, studied by NASA to put a several-meter asteroid into cis-lunar orbit for astronauts and robots to practice on [15]. For substantially lower cost, and potentially even greater relevance for its accesses to asteroidal 'gravity', although less persistent (1-year) and limited to a ~20-cm 'patch of regolith', AOSAT+ will provide a proxy opportunity in a semi-automated laboratory.

In this paper, we introduce the AOSAT+ mission concept. AOSAT+ is a 12U CubeSat-based centrifuge planetary science laboratory that will be located in Low Earth Orbit (LEO). The centrifuge will spin at up to 1.1 RPM to generate artificial milligravity that would be experienced on asteroids 22 km and smaller (Table 1).

**Table 1: Surface Gravity Simulated by AOSAT+**

| Asteroid | Diameter (km) | Gravity (m/s$^2$) | RPM |
|---|---|---|---|
| Didymos-B | 0.16 | $5·10^{-5}$ | 0.14 |
| Bennu-Ryugu | 0.5-0.9 | $9·10^{-5}$ to $2·10^{-4}$ | 0.18-0.25 |
| Deimos-Phobos | 12–22 | $3·10^{-3}$ | 1.1 |

AOSAT+ a science investigation whose mission objective is to make substantial, quantitative inroads into the underlying regolith physics of asteroids, comets and small moons. It will acquire *direct* knowledge about surface strength and regolith response/compliance to interrogation by robotic platforms; granular dynamics in ultra-low (but non-zero) directional gravity fields; gas entrainment as volatiles interact with and permeate through loose regolith; ballistic penetration of particles striking the surface; material interactions between asteroid regolith and spacesuit fabrics and glass/metal surfaces, the effects of vibrations and gas jets on dust mitigation, and (with the SEO) the controlled mobility of a robotic interactions with asteroid regolith materials.

In the following section we present an updated background of centrifuge concepts being considered and being planned for launch in space. This is followed by presentation of the spacecraft overview in Section 3, Concept of Operations in Section 4 and overview of the science instruments in Section 5, followed by discussions and conclusions.

## 2. BACKGROUND AND RELATED WORK

Simulating asteroid surface conditions remains a formidable challenge. The challenge comes from simulating the low-gravity conditions. Figures 2 present some conventional methods to simulate low-gravity conditions on earth. These include parabolic flight, use of neutral buoyancy within large water tanks and drop towers. In a parabolic flight, an aircraft is maneuvered to create brief periods of micro-gravity conditions last 10-20 seconds [16] Neutral buoyancy methods suspend objects in water, with buoyant attachments that compensates for the objects mass [17]. Finally, drop towers contain a chamber that fall for 5-10 seconds enabling the contents of the chamber to briefly experience microgravity conditions [18]. These conventional methods simulate low-gravity conditions for too brief a time period or impose simulation artifacts that prevent correlation with real asteroid surface conditions.

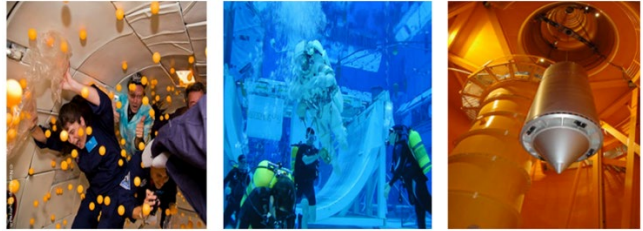

**Figure 2. Methods to simulate low-gravity conditions include use of parabolic flight (left), neutral buoyancy in large water tanks (center) and use of drop towers (right).**

A promising solution to simulating low-gravity conditions is using a centrifuge operating in Low Earth Orbit [19]. The centrifuge consists of a mass *m*, spinning at a radius *r*, at an angular velocity ω, which create a centrifugal force of magnitude $F_c = m\,r\,\omega^2$. The concept of a space centrifuge is not new and has been a popular topic of science fiction. However, we have yet to see a habitation centrifuge or centrifuge science laboratory operate in space. Our focus is to create a centrifuge science laboratory to simulate the surface conditions and the physics of small bodies.

We propose to do this by creating a small testbed in space. Several centrifuges onboard the ISS already do various kinds of fundamental physical and biological science [20], and while asteroid gravity is typically orders of magnitude smaller than existing capabilities, one should consider developing an asteroid research centrifuge onboard the ISS. The most critical problem is that the required experiments would have to contend with the relatively strong vibrations from spacecraft pumps and fans, typically ~0.01 cm/s$^2$ with peaks several times that; also there are turbulent forces within the pressurized habitable environment.

Ambient vibrations are comparable to the gravity of the smallest of our experiments ($10^{-5}$ g) and would greatly limit the quality of our study of asteroid geophysics. The injection of random energy would fluidize unconsolidated materials and influence geological stability, processes that we plan to study [21]. The size of the spinning centrifuge, that would itself have to be in vacuum, is far too large for the vibration isolation facility onboard the ISS [21], presenting a major logistical challenge. Furthermore, a large spinning centrifuge adds operational challenges to the ISS, as the angular momentum added (kilograms of centrifuging material) needs to be nulled by the ISS GNC.

A free-flying centrifuge is simpler and more cost-effective to design, will last for well over a year in Low Earth Orbit, and provides excellent downlink, and will attain cleaner milli-gravity conditions than onboard the ISS, and will do so under asteroid-like vacuum conditions. In our earlier work, we have proposed utilizing a 3U CubeSat to test the concept. CubeSats are emerging as low-cost platform to perform space science and technology research. They offer the possibility of short development times, wide use of Commercial-Off-the-Shelf Technologies (COTS), frequent launches and training of students.

Our first CubeSat science laboratory mission is called Asteroid Origins Satellite-1 (AOSAT-1) [11-14]. The spacecraft, a 3U, 34 cm × 10 cm × 10 cm (size of a loaf of bread) will contain a science chamber that takes two-thirds of the spacecraft volume and contain crushed meteorite (Figure 3). One third of the spacecraft contains the spacecraft electronics, Guidance Navigation and Control (GNC), communications electronics and the power system. The spacecraft produces artificial gravity by spinning at upto 1 RPM, using magnetorquers and a reaction-wheel. This is sufficient to simulate the gravitational forces experienced on a sub 1 km asteroid. A major advantage of a centrifuge science laboratory such as AOSAT 1 is that it can simulate asteroid surface conditions without having to go to an asteroid, which remains a long and expensive endeavor. These centrifuges can help use prepare for future mission by testing new technologies under asteroid conditions.

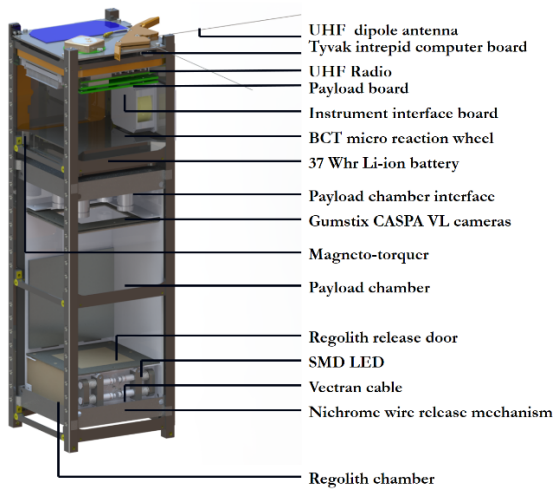

**Figure 3: AOSAT 1 is On-Orbit 3U CubeSat Centrifuge Laboratory that is going to be launched in 2019.**

### 3. SPACECRAFT OVERVIEW

AOSAT+ (Figure 4, 5 and Table 2) is a 12U, 24 kg CubeSat that will operate as on-orbit centrifuge science laboratory to simulate the surface conditions of small-bodies such as Didymos-B, Bennu, Ryugu and Phobos/Deimos. AOSAT+ will operate as a centrifuge by spinning about its short axis. The will produce a *centrifugal force* to simulate asteroid gravity. AOSAT+ will utilize a suite of onboard instruments to characterize the dynamics and behavior of crushed meteor particles under asteroid surface gravity conditions relevant to planetary science, asteroid robotics and ISRU.

The 12U bus is developed by the Space Dynamics Laboratory (SDL) and has already been used on several DoD missions. The spacecraft will be powered using two large body-mounted deployable eHawk solar panels containing triple junction cells. The MMA eHawk (TRL 9) will avoid gimbaling which a provides major simplification for the operation of the spacecraft. The MMA eHawk has been widely selected for interplanetary CubeSats on the SLS-EM1 mission and on JPL's MarCO mission to Mars. The system will charge a bank of 144 Whr Yardney Lithium Ion batteries (TRL 9). Depth of discharge will not exceed 20 % to maximize battery capacity and life. Preliminary power subsystem design suggests a 25 % margin (Table 2), sufficient for meeting the needs of the spacecraft.

This unique 12U CubeSat design provides robust mass margin of 25%, volume margins of 44%, power margin of 25% and that includes 2 to 25% contingency (Table 2). The spacecraft uses a SDL built Single Board Computer containing Cobham Leon 3FT rad-hardened processor (TRL 9) operating at 125 Mhz coupled with a rad-tolerant Microsemi FPGA, 256 Mbyte of SDRAM and 16 GB solid state drive. The SBC board interfaces to the Mission Unique Card (MUC) via SpaceWire. The MUC hosts all the science instruments. The SBC is rad-tolerant thus enabling robust handling of Single Event Upsets (SEUs) and Random Bit Flips (RBFs). The proposed computer board has been successfully combined with the proposed Innoflight SCR-101 S-band Radio (TRL 8).

The SDL single board computer while being extremely power efficient and radiation tolerant, lacks the required computational capability to perform the required data processing from the science instruments, particularly for running vision and other data processing algorithms. Hence these functions are performed using the 4 Raspberry PI 3 Compute Modules (TRL 9) that acts as a controller for each instrument and are connected to the MUC card.

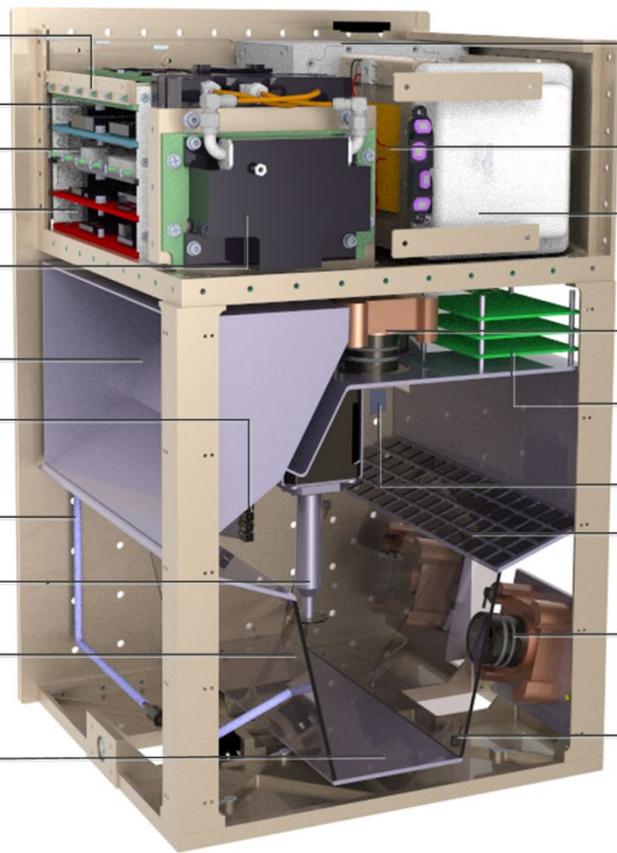

**Figure 4: AOSAT+ Internal Layout**

The spacecraft will use the Innoflight SCR-101 S-band Radio capable of 1.5 Mbps data rate. The system is designed to send/receive at least 8-bit tone data with a 6-dB margin even if the spacecraft is spinning. SDL will be providing spacecraft tracking and communication services using the US government run MC3 communications network. The on-board Attitude Determination Control System (ADCS) consists of the Blue Canyon Technologies XACT-50 that combines a suite of sensors such as a star tracker and sun sensor to perform space inertial measurements. For attitude control the system contains 3-axis reaction wheels.

In addition, the spacecraft is equipped with a 100 mNms reaction wheel in the axis of spin for a total of 150 mNms. In addition, magne-torquers will be used to periodically desaturate the reaction wheels. This provides a total space attitude determination and control system package. The on-board attitude control solution is expected to provide accurate pointing capability of 1 degree sufficient for achieving 1.5 Mbps of downlink using the baselined Innoflight S-band transceiver and MC3 Ground Network.

**Table 2: Mass, Volume and Power Budget**

| System | Mass (kg) | Volume (cm$^3$) | Mass & Vol. Contingency | Avg. Power (W) |
|---|---|---|---|---|
| Communications | 0.3 | 55 | 2% | 4.4 |
| Onboard CPUs | 0.2 | 120 | 2% | 3.2 |
| Instruments | 8.2 | 3300 | 14.5% | 16 |
| Power Conv. | 2.1 | 1400 | 2% | 2.8 |
| Navigation | 1.4 | 350 | 2% | 4.2 |
| Structure | 4.7 | 960 | 25% | - |
| Thermal | 0.3 | 820 | 10% | 0.5 |
| Total | 18.0 | 11200 | - | 32 |
| **Margin** | **25%** | **44%** | **-** | **25%** |

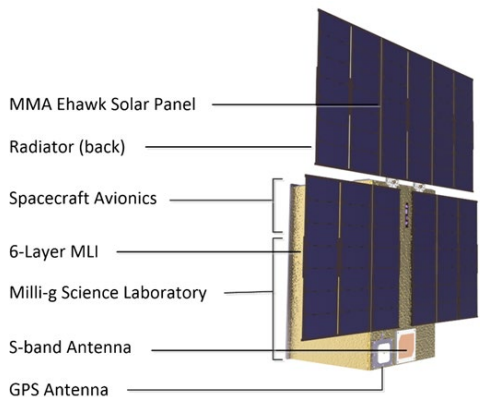

**Figure 5. AOSAT+ External Layout**

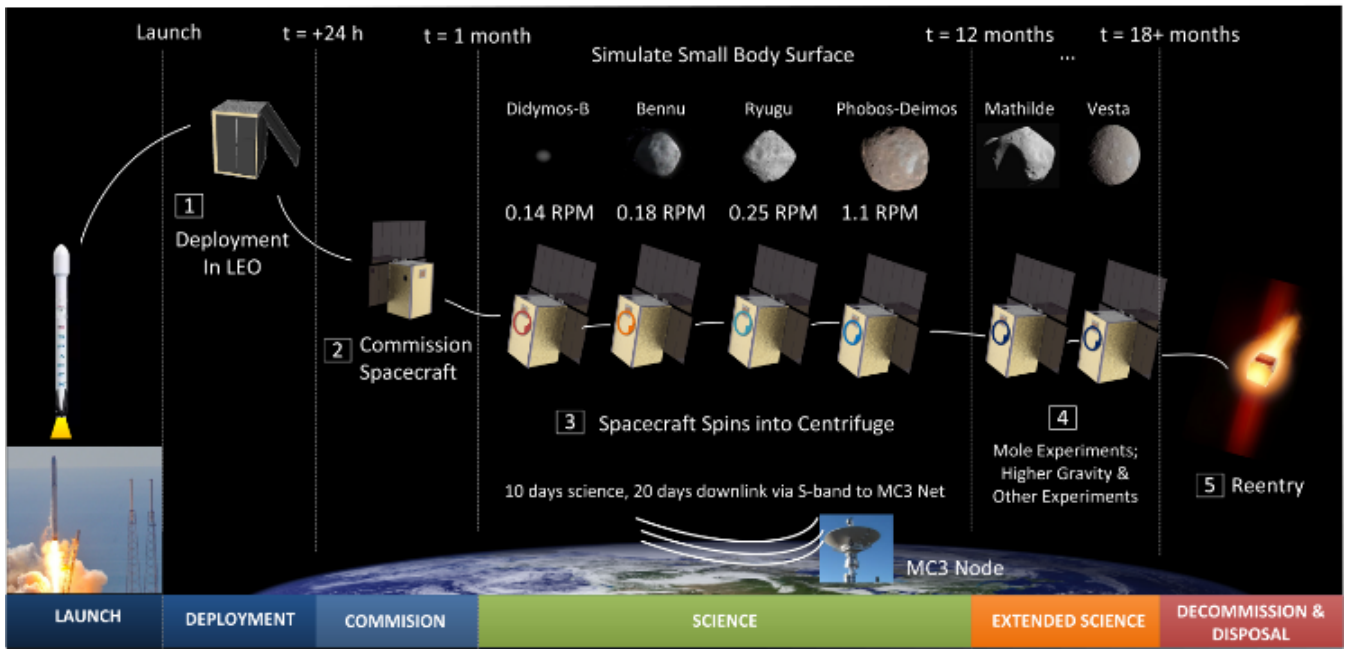

**Figure 6: AOSAT+ Concept of Operations**

## 4. MISSION CONCEPT OF OPERATIONS

The Concept of Operations is shown in **Figure 6**. The spacecraft will be ejected from a Planetary System's Corp Canisterized Satellite Dispenser (CSD) at a velocity of 1 m/s. As the solar panels are spring deployed, they will automatically deploy open once exiting the CSD. After 30 seconds, the spacecraft computer will boot and perform major systems check. Upon health-checks on the battery and ADCS, the spacecraft will execute a detumble procedure to get into 'home' position. This will begin the formal commissioning of the spacecraft over the first-month and preparatory science experiments will be conducted to determine the health of the instruments.

The Science Investigation will consist of twelve 10-day sets of experiments, each set generating ~20 GB of compressed data, to be followed by 20-days of post-experiment downlink and analysis on the ground. Only limited data can be transmitted while the laboratory is spinning, namely high-level status of the experiment, and thus we make use of simple onboard autonomy to sense when an experiment is finished, to shut down its data-stream and prepare for the next experiment. The science objectives are to use Ecliptic CMOS cameras and tactile sensors and accelerometers to monitor granular behavior and interactions as the spacecraft maintains a set angular velocity and the regolith responds to vibrations, a penetrometer, a nitrogen Gas Jet, and a bead deployer.

The dominant data volume consists of high resolution jpegs and HDTV movies. The jpegs and tactile sensor data will be sent first, while the camera controller compresses the HDTV movie files to be sent as they are generated. The schedule in Table E8 will be followed resulting in 330 days of science operations during the first year resulting in a total of 300 GB of data downlinked to ground and excludes communications overhead. An extended mission will be for 180-days and conclude with two months of robotics experiments with the 'mole', followed by higher gravity experiments to simulate the surface regolith of Mathilde and Vesta.

## 5. SCIENCE LABORATORY

The AOSAT+ lab chamber is housed in the lower two-thirds of the spacecraft away from the flight avionics and maximizes the centripetal radius to attain maximal-g for lowest spacecraft spin (angular velocity). The chamber is shown in Figure 7. Regolith (black) is shown in its stowed configuration; the door drops and the regolith comes out, aided by centrifugal force and vibrations. Shown are the spacesuit material coupons, the tactile skin, optical systems (includes cameras, LEDs and laser sheets), and actuators (including vibrators, Gas Jet, bead deployer). The mole (SEO) will be mounted inside the regolith stowage chamber, for deployment after Baseline EOM.

**Optical Camera System.** AOSAT+ will have 4 framing cameras for recording science experiments: two stereo pairs, and two that image the glass wall. The camera interface consists of a ruggedized COTS camera-lens sensor head assembly to a common image-processing/avionics interface board similar to flown units**.** Sensor data is passed to the Image Processing Board (IPB) with non-volatile program memory and sequence storage. The IPB also controls the short-circuit protected power and incorporates watchdog timers and current monitoring to ensure recovery from SEUs. The IPB enables real-time processing functions (windowing, compression, binning) that will be used to reduce the data footprint of each experiment

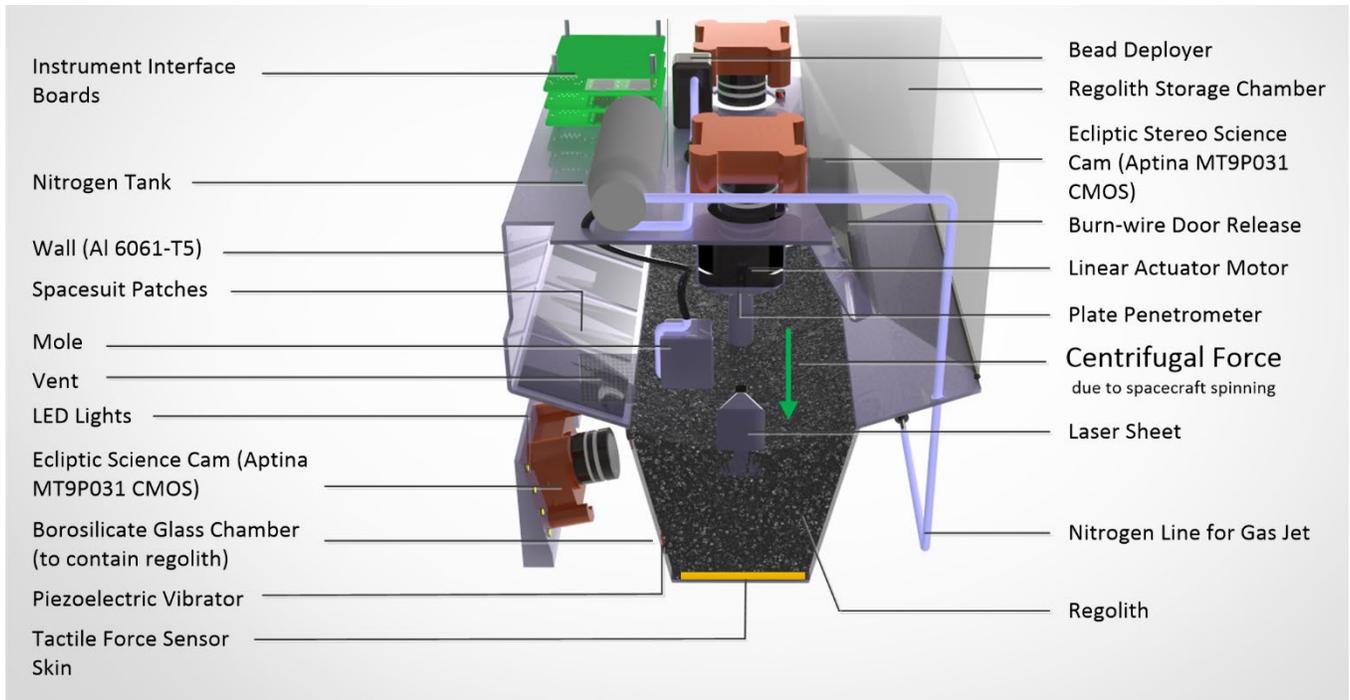

**Figure 7: AOSAT+ Centrifuge Laboratory**

**Bead Deployer.** To study higher energy ballistic interactions between particles, we use a 'pinball deployer' that can launch ten 2.5 mm aluminum spheres at a velocity of 10 to 100 cm/s to strike the regolith and form a crater. The baselined design of a pneumatic bead deployment system utilizes controlled bursts of low-pressure $N_2$ as a carrier for each bead. Control is affected through an in-line latching valve that opens and closes via an electrical control signal.

**Gas Jet.** A 17 $cm^3$ tank of $N_2$ gas at 0.006 bar starting pressure is used to emit controlled pulses through an in-line latching valve to provide repeated inert gas ($N_2$) interaction with regolith. Nitrogen is being used as the driver for OSIRIS-REx sample return [6] and is being planned for use in the CAESAR comet nucleus sample return [22].

**Vibrators.** Vibrations provide simple, solid state regolith actuation. A cell phone vibration, transmitted throughout the volume of a 12U CubeSat rotating at 1 RPM, is sufficient to send all the regolith flying out of its bed. There will be 12 vibrators in all, tunable to much lower energies and capable of small haptic pulses.

**Tactile Force Sensors**. Flight units, consisting of a 142-taxel 'skin' lining the regolith bed, 13 on the force bearing head, and 108 on the Mole robot, will be developed [23]. Each skin will be capable of measuring normal force and 2D shear force at 100 Hz at a 1 cm spatial resolution. Each taxel will have a sensitivity of 0.1 kPa, range of 0.1 to 1000 kPa.

**The Mole.** The Mole, a 50 g lander-type system, has one technology objective, namely *to demonstrate one or more modes of controlled mobility on low-gravity surfaces [33-35, 38]*. In terms of delta-v, the cost of hopping or flying over an asteroid is significantly low; however, mobility of any kind in a milligravity environment requires very precise control. Demonstrating this mobility inside of AOSAT+ requires precise control in a very small package. Our proposed concept involves taking our earlier SunCube 1F FemtoSat concept that consists of 3 cm × 3 cm × 3 cm LEO spacecraft platform [24] and adapting it towards low-gravity surface exploration. The SunCube FemtoSat contains a stack of electronics containing IMU, CPU (32-bit ARM processor), lithium-ion-battery, 3-megapixel camera, and a board for controlling miniature servos.

The SunCube has options for wireless radio, however that poses integration challenges in such a small volume. A wireless experiment could be used to pilot a small robot as it 'explores' the patch of regolith that is inside of AOSAT+. It could try to burrow, hop, orient itself, inertially free from cables and tethers. The challenges in power and communications makes us consider instead the logically much simpler, and lower risk, option of tethering the Mole to the spacecraft, to providing power to recharge the battery. The downside is that a tether will apply a perceptible force to the Mole.

For mobility we shall consider mechanical and gas-driven approaches. For instance, with the addition of a thin flexible hose we could pipe $N_2$ gas that would allow for experiments to directly explore the use of blown gas as a propellant to enable long-duration missions on the surfaces of asteroids, someday using in-situ volatile resources [36-37]. One advantage to this approach is that the valve design would borrow heavily from the bead deployer and gas-jet and be developed in parallel with those. The gas would be blown out of a small nozzle, simulating an onboard propulsion system. Servos onboard would adjust the pointing of the nozzle to see how the system (Mole embedded in regolith) reacts. The experiments will be used to develop basis-behaviors will be used to perform autonomous navigation on a maze-like debris field-filled asteroid surface [39].

**Plate Penetrometer.** The strength of a regolith surface will be measured using a force-bearing plate with a linear actuator rod and bi-directional force sensor. The resisting force will be recorded at 100 Hz as the actuator extends from above the regolith to within $d_{max}\sim$1cm (size of the largest particles) from the bottom. The design is inspired by the slow mode of penetration of comet penetrometers [25].

## 7. PLANNED EXPERIMENTS

After the spacecraft undergoes commissioning for the first month, the Threshold Mission proceeds for 9 months, with each 30 days defining one set of experiments. There are 3 sets of experiments for each artificial gravity (Table 1) during a 10-day session of lab work, followed by 20 days of downlink, producing >20 GB of data. Three subsets of investigations take place: 3 days on *Bearing Strength*, 3 days on *Gas Jets*, and 3 days on *Granular Behavior*, and 1-day margin. These sets will be conducted as automated sequences recalibrated after every session and repeated for the various 'asteroids' (spin rates). *Material Interactions + Dust Behavior* are studied during all of these experiments. The Baseline Investigation adds 2 further months of experiments (completing the 1-year science mission) with *Ballistics* and *Brazil Nuts* depending on the action of the ball deployer, one-set at Didymos-B gravity and one set at Ryugu gravity, as both of these targets (by NASA DART mission and JAXA Hayabusa ) will experience artificial impacts.

Once the centrifugal force begins, the regolith settling timescale is in minutes (Figure 8) depending on particle dynamics; we allow 1 day of schedule. The spin can be increased and then decreased, creating an over-steepened pile for some experiments. The tactile force sensors and accelerometers will confirm cessation of particle movement. In Phase A we shall begin development and testing of algorithms for automating the setup and the 'playlist' of experiments. Automation and other aspects of spacecraft internal operations such as data compression, can be modified and optimized during the mission, with the levels of human interaction decreasing and the data quality and performance increasing as the mission progresses, with ongoing validation.

**Bearing Strength.** To test the solidity of regolith we will press a 2-cm diameter disk into and through the regolith using a linear actuator rod. A tactile skin will record forces at 100 Hz as the actuator extends from above the regolith to within ~1 cm (size of the largest grains) from the bottom. Force sensing (pressure and x-y shear) at the base of the pool allows bearing-stress profile to be obtained. Images of regolith displacement will be taken by top and side cameras.

**Granular Behavior.** Every simulated asteroid gravity will include an angle of repose measurement and a landslide experiment, plus other experiments as time permits, vibrating or de-spinning the granule bed to generate new physical conditions for analyzing granular behavior. These experiments will be modeled numerically (Figure 8), matching the high definition images and movies and force sensor readings to develop particle statistics on coefficient of restitution, tangential friction, and rolling friction [40]. These models will be validated visually by matching image data to ray-traced simulation visualizations, using open-source codes POV-Ray and ParaView.

**Gas Jets.** To study how regolith is affected by gas in the near-surfaces of volatile-rich asteroids, comets, and planetesimals in the nebula, and in sampling approaches e.g. OSIRIS-REx, a 17 cm$^3$ tank of N$_2$ at 0.006 bar starting pressure will be piped below the regolith bed as shown in Figure 7. A valve will release up to 30 pulses. The gas moves through the regolith (or entrains it) and escapes in a mesh of 0.3-mm vent holes in the regolith stowage area.

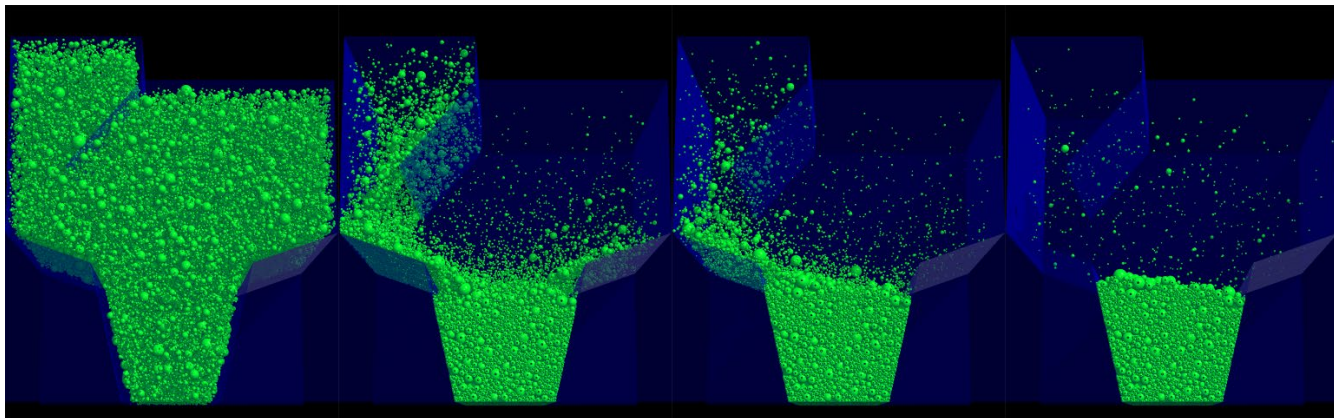

**Figure 8**. Simulation of AOSAT+ payload chamber spinning at 0.25 RPM (Schwartz et al. 2014 [28]). Simulation begins with regolith dispersed as a granular-gas (*left frame*). After a few minutes of collisional dissipation and under influence of the centrifugal acceleration in the chamber, grains begin to coagulate, generating a granular-liquid phase-transition, and funnel into the pool. Frames show progression in time from left to right. Size distribution $N(d) \sim d^{-3}$ ranges from 1 mm to 1 cm (~10$^5$ particles). Coefficient of restitution is 0.8, although it may be closer to 1 [29]. Recording this behavior and matching simulations to experiments will provide unique model validation and essential constraints on regolith parameters.

Valves can leak in dusty environments; so if this happens we will purge the tank before the remaining experiments. Additionally, gas may prove to be an effective way of clearing dust.

**Ballistics.** Ten 2.5 mm aluminum spheres will be launched at 10-100 cm/s into the regolith bed by the gas-driven mechanism shown in Figure 7, using the same $N_2$ tank, with the gas vented to space so as not to influence the regolith. The impact of these spheres, meant to simulate fallback of ejecta at around the escape velocities of the simulated asteroids, will change the regolith bed, producing dust and adding metallic sphere and so will be done after the threshold investigation is complete. Measurements will include depth of penetration and transient crater size. Laser imaging will give the mass-velocity distribution of ejecta. These data will be compared with *pkdgrav* simulations that match the crater opening and timescale of resettling, providing strong constraints on the rate-dependent coefficients of restitution and other parameters. Transient forces that are resolved at the base of the sheet will provide information on velocity attenuation and seismology in a pile of regolith, of critical importance to cratering mechanics and asteroid deflection. Schwartz led a study successfully matching the results of similar-speed experiments into similar-size particles in Earth gravity, by Makabe & Yano (2008) [30], by analyzing the sampling mechanisms of the Hayabusa [31] and Hayabusa 2 [5] missions.

**Brazil Nuts.** The ballistic impacts will add aluminum spheres to the regolith. Once the impacts are finished, a Brazil Nut Effect sequence is initiated, to vibrate the bed and observe with the side cameras any size-sorting or segregation – whether the spheres or other grains aggregate, rise to the surface [32], get trapped below or in corners, or accrete/repel dust.

## 7. DISCUSSION

Small body exploration is in a bootstrapping phase where we need to know more to discover more. Until then, missions to asteroids and comets and their adventures on the surface are full of risk. So OSIRIS-REx, a mission whose primary objective is sample return, will spend nearly two years obtaining spectacular remote sensing data, then spend only a few seconds minimally interacting with the surface to collect a sample (using $N_2$ gas entrainment).

Further physical interaction is avoided once the science goal is met. Hayabusa-2 plans to deploy dropped landers (MASCOT [26] and Minerva II) and a small carry-on impactor [27] but these are geared for surface and subsurface composition discovery while the information on grain-scale regolith mechanics will be limited. The Lucy and Psyche missions will provide unprecedented views of diverse asteroids, but no exploration of the surface. AOSAT+ can fill in some of these missing pieces, bringing a Bennu-like testbed and other proxies to LEO, allowing many relevant experiments whose *mission-enabling data* could not be obtained in any other way, and that could help make sense of mission discoveries since NEAR and the ongoing discoveries at Bennu and Ryugu

## 8. CONCLUSIONS

AOSAT+ is an on-orbit centrifuge laboratory that will be used to perform basic planetary science experiments in a realistic asteroid gravity and surface environment. The mission will help to characterize the physics, interaction behavior, and near-surface mechanics of Didymos-B, Bennu, Ryugu, Deimos and Phobos by creating a research proxy in Low-Earth Orbit, supported by ground-based meteorite analysis and 3D computer modeling. Knowledge gained will help validate proposals and mechanisms for hazardous asteroid mitigation, mining, and advanced exploration of small planetary bodies. This approach to simulate low-gravity conditions provides a persistent link to off-world environments. Using such laboratories, it is possible to simulate alien environments (different gravity, atmospheric pressure, electrical conditions and so on) and test hypotheses for unknown or poorly understood planetary surface processes; this, in turn, may be used to validate computer models in order to develop advanced simulation proxies for science, exploration, mining, habitation, and hazardous asteroid deflection. By recreating alien surface environments, we can test and validate robotic landing technology and eventually human adaptation to these environments and broaden our understanding and prove the feasibility of risky off-world surface exploration techniques before going to these locations bodies.

## ACKNOWLEDGEMENTS

The authors would like to gratefully acknowledge the AOSAT+ Science CoIs, Drs. Figueroa, Hartzell and Santos, the DSRI team, Chris Shinohara and Patrick Kerr, the SpaceTREx engineering team including Aman Chandra, Leonard Dean Vance, Himangshu Kalita and Ravi Nallapu, for helping to develop the mission concept.

## BIOGRAPHY

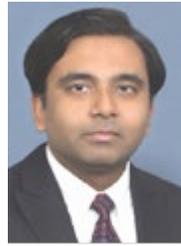

***Jekan Thangavelautham*** *has a background in aerospace engineering from the University of Toronto. He worked on Canadarm, Canadarm 2 and the DARPA Orbital Express missions at MDA Space Missions. Jekan obtained his Ph.D. in space robotics at the University of Toronto Institute for Aerospace Studies (UTIAS) and did his postdoctoral training at MIT's Field and Space Robotics Laboratory (FSRL). Jekan Thanga is an assistant professor and heads the Space and Terrestrial Robotic Exploration (SpaceTREx) Laboratory at the University of Arizona. He is the Engineering PI on the AOSAT I CubeSat Centrifuge mission and is a Co-Investigator on SWIMSat, an Airforce CubeSat mission concept to monitor space threats*

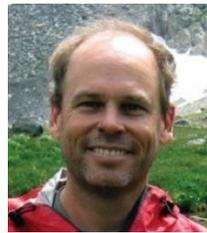

***Erik Asphaug*** *graduated in Mathematics and English from Rice University, taught high school for three years in Tucson, and obtained his PhD in Planetary Sciences from the University of Arizona in 1993. After a postdoc at NASA Ames he was hired and helped found the Planetary Sciences program at UC Santa Cruz. His signature research has been the geology of comets and asteroids, the physics of collisions, and planet formation. Since being at University of Arizona's Lunar and Planetary Laboratory he has been able to combine his deep interests in theory, laboratory work, and space exploration, for example as the Science PI of AOSAT I, a microgravity research centrifuge selected by NASA for launch, whose goal is to answer primary questions about accretion physics and asteroid regolith environments.*